\documentclass[fleqn]{article}
\usepackage{graphics}
\begin{document}
\newcommand{\keywords}{general relativity, numerical relativity, black holes} 
\newcommand{\PACS}{04.25.Dm, 95.30.Sf, 97.60.Lf}
\newcommand{\shorttitle}
{B. Br\"ugmann, Numerical Relativity} 
\title{Numerical Relativity in 3+1 Dimensions}
\author{
\\
  {Bernd\ Br\"ugmann}
\\ \\
  \small Max-Planck-Institut f{\"u}r Gravitationsphysik 
\\
  \small (Albert-Einstein-Institut) 
\\ 
  \small Am M\"{u}hlenberg 1, 14476 Golm, Germany
\\ 
  \small {\tt bruegman@aei-potsdam.mpg.de}
} 
\date{\small December 2, 1999; AEI-1999-40}
\maketitle
\begin{abstract}
  Numerical relativity is finally approaching a state where the
  evolution of rather general (3+1)-dimensional data sets can be
  computed in order to solve the Einstein equations.  After a general
  introduction, three topics of current interest are briefly reviewed:
  binary black hole mergers, the evolution of strong gravitational
  waves, and shift conditions for neutron star binaries.
\end{abstract}

\section{Introduction}
\label{introduction}

In this short review I describe work concerned with one of the central
issues of numerical relativity, the solution of the two body evolution
problem of general relativity.  After a short introduction to
(3+1)-dimensional numerical relativity, I briefly discuss recent
pro\-gress on binary black hole mergers, the evolution of strong
gravitational waves, and shift conditions for neutron star binaries.

As opposed to Newtonian theory, where the Kepler ellipses provide an
astrophysically relevant example for the analytic solution of the two
body problem, in Einsteinian gravity there are no corresponding exact
solutions. The failure of Einstein's theory to lead to stable orbits
is due to the fact that in general two orbiting bodies will emit
gravitational waves that carry away energy and momentum from the
system, leading to an inspiral.  But, of course, this ``leak'' is not
considered to be detrimental. Gravitational waves are one of the most
interesting new phenomena introduced by general relativity that will
open a new window into the universe through gravitational wave
astronomy, e.g.\ \cite{Schutz99}.
  
The evolution of a two body gravitational system, for example a binary
black hole system (which can be constructed as a vacuum system and
avoids additional complication due to matter sources), can be divided
into at least three phases.  For sufficiently large separation of the
two black holes there is a slow inspiral phase with many orbits,
followed by a very brief violent merger phase that leads to a single,
distorted black hole that after a short ring-down phase settles down
to a final stationary black hole. For the initial and final phase
rather well understood approximation schemes are available, i.e.
post-newtonian calculations for the slow inspiral of two point masses
(e.g.\ \cite{Damour97,Blanchet97}) and the close limit approximation
for the ring-down of a single distorted black hole(e.g.\ 
\cite{Pullin98a}). For a full treatment of the strongly non-linear,
fully general relativistic phase one has to turn to computer
simulations to obtain (again approximate) numerical answers.

Each phase leads to a characteristic gravitational wave signal.  At
this time several gravitational wave detectors are being built
world-wide that should for the first time make the direct measurement
of gravitational waves possible. The prediction and analysis of future
signals is the main motivation for studies of binary systems in
numerical relativity.  While certainly the primary motivation, let me
add that even if it had no direct, in the near future measurable
observational consequences, we should solve a basic problem like the
two body problem of general relativity.

The article is organized as follows.  In Sec.\ \ref{history}, a
brief history of black hole evolutions in numerical relativity in 2+1
(axisymmetry) and 3+1 dimensions is given. In Sec.\ \ref{cauchy},
the evolution problem of numerical relativity is introduced in its 3+1
form, leading to three main issues: initial data, evolution, analysis.
Initial data is computed on a three-dimensional hypersurface, which is
evolved in time, and at various times analysis like identifying the
black hole horizons and gravitational wave extraction is carried out.
The coordinate problem of numerical relativity is emphasized with the
choice of slicing function, the lapse, as an example. The skeleton of
a typical black hole evolution is discussed. In Sec.\ \ref{cactus},
the ``Cactus'' code, a computational infrastructure for numerical
relativity and relativistic astrophysics, is described.

After this general introduction, we discuss several examples for
recent pro\-gress in (3+1)-dimensional numerical relativity.  In Sec.\
\ref{current}, current (summer of 1999) binary black hole simulations
are presented. The holes start out close to each other and evolve
through a plunge rather than an orbit (a grazing collision).
Achievable evolution time is now about $30M$ ($M$ the mass of the
final black hole), which for the first time allows the extraction of
wave forms. In Sec.\ \ref{waves}, a discussion of strong wave
evolutions is included because strong waves play a role in black hole
mergers and these studies provided the proving ground for a new
evolution scheme discussed in Sec.\ \ref{adm}. In Sec.\
\ref{shift}, the minimal distortion shift condition is described.
Lapse and shift specify the coordinate gauge, and in all simulations
mentioned so far the shift has been zero, but for systems with
rotation a shift condition will be essential. The example presented is
minimal distortion shift for a binary neutron star system, which for
this purpose is simpler than black holes because there are no special
inner boundaries.

Sec.\ \ref{conclusion} concludes this brief review, pointing out
again those issues and techniques that will be important for the
numerical simulation of binary black holes for several orbits lasting
for $100$--$1000M$ with results that are relevant for gravitational
wave astronomy.

\section{History of black hole simulations in numerical relativity}
\label{history}

In this section I endeavor to give a necessarily very short but in its
highlights complete exposition of the literature on numerical black
hole evolutions, concentrating mostly on work that implements the
complete black hole evolution problem (data, evolution, analysis) for
the full Einstein equations in vacuum. Matter occurs in a few places
but only as a means to form black holes. Clearly, there is a large
and important body of work concerned with all the different, separate
aspects of and methods for black hole evolutions as outlined in
Sec.\ \ref{cauchy}. Still, this allows us to sketch the history of
the field.

\subsection{2+1 dimensions}
\label{2+1}

After some early attempts \cite{Hahn64}, it was the work by Smarr
\cite{Smarr75,Smarr77} and Eppley \cite{Eppley75} on the head-on collision of
two equal mass Misner black holes \cite{Misner63} which basically founded
the field of numerical relativity as a subject of computational
physics. Axisymmetric head-on collisions allow significant
savings in computational cost when formulated with two spatial and one 
time coordinate (2+1 dimensions), although this excludes the possibility of 
orbiting black holes and radiation of angular momentum. Many of the
key techniques that are still in use today stem from that period of
the sixties and seventies (see \cite{York79} for the definitive
review). 

The beginning of the nineties saw a surge of activity when more powerful
computers, improved codes and methods allowed significant advances.
The axisymmetric collision of black holes, either formed by particles
\cite{Shapiro92a} or implemented as Misner data
\cite{Anninos93b,Anninos94b}, was repeated complete with horizon
finding and wave extraction. It is remarkable how the crude results
for the wave emission of \cite{Smarr75,Smarr77} where confirmed in
\cite{Anninos93b}. Another highlight is certainly the numerical
computation of the ``pair of pants'' picture for a black hole merger
\cite{Matzner95a}, which was a result of the US Binary Black Hole
Grand Challenge Alliance. Head-on collisions in axisymmetry continue
to improve, see the recent work on unequal mass configurations
\cite{Anninos98a,Brandt99}.

Another interesting system in axisymmetry are rotating black holes. In
\cite{Hughes94a}, particles collapse to form a black hole with
rotation and toroidal event horizon. In \cite{Brandt94c}, a Kerr
black hole distorted by a gravitational wave is evolved. Matter plus
rotating black hole systems are also studied in \cite{Brandt97e,Brandt98}.

A traditional topic in numerical black hole studies is that of black
hole formation, of which I want to mention only the following recent
references that are of relevance to this article.
The formation of naked singularities was examined in \cite{Shapiro91}.
Furthermore, in \cite{Abrahams92b} the collapse of gravitational waves
to a black hole is demonstrated.  A surprise was that even
(1+1)-dimensional, spherically symmetric black hole systems are far
from being trivial, as the rich set of critical phenomena discovered
by Choptuik \cite{Choptuik93} showed (see e.g.\ \cite{Gundlach97d} for
a review).  In 2+1 dimensions, the only critical collapse studies so
far are those of \cite{Abrahams93a}.

\subsection{3+1 dimensions}
\label{3+1}

Numerical relativity of black holes in 3+1 dimensions was initiated in
1995 with evolutions of a Schwarzschild black hole with singularity
avoiding slicing on a Cartesian grid \cite{Anninos94c}. Achieved run
time is about $30M$. At the same time the first (3+1)-dimensional wave
simulations were carried out \cite{Shibata95,Anninos94d}. In Sec.\
\ref{waves}, I comment on the collapse of non-axisymmetric waves to a
black hole \cite{Alcubierre99b}. 

Returning to our main topic, the evolution of a Schwarzschild black
hole with a non-vanishing shift vector was studied in \cite{Daues96a},
compare Sec.\ \ref{shift}.  In \cite{Bruegmann96}, adaptive mesh
refinement techniques, made famous in numerical relativity by
\cite{Choptuik93}, were applied for the first time to 3+1 relativity,
also for a Schwarzschild black hole.  By now, evolutions for the
Schwarzschild spacetime are standard code test, e.g.\ \cite{Bona98b}.
Further studies of single black holes include the distorted black
holes in \cite{Camarda97a,Camarda97c,Allen98a}, which provided the
first detailed tests of wave extraction in 3+1 dimensions. The Black
Hole Grand Challenge Alliance performed the longest, stable evolution
of a single black hole so far, reaching about $60M$ for a standard
Cauchy evolution with black hole excision and a boosted black hole
\cite{Cook97a}, and essentially achieved complete stability ($>
60000M$) with a characteristic evolution code, which is tailored to
the one black hole problem but can also treat small distortions, and
for the first time a black hole that moves across the grid
\cite{Gomez97b,Gomez97a,Gomez98a}.

Binary black hole evolutions are pushing the limits of what is
currently possible. Some results for the evolution of the axisymmetric
Misner data set with the 3+1 code of \cite{Anninos94c} with
singularity avoiding slicing are reported in \cite{Anninos96c}. The
first true (3+1)-dimensional binary black hole evolutions, the grazing 
collision of nearby spinning and moving black holes, was performed in
\cite{Bruegmann97}. This sets the stage for the recent binary black
hole simulations of Sec.\ \ref{current}, but first we want to discuss
some of the basic issues in numerical relativity.

\section{Anatomy of a numerical relativity simulation}
\label{cauchy}

\subsection{3+1 formulation}
\label{adm}

The Arnowitt-Deser-Misner (ADM) equations \cite{Arnowitt62,York79} are
one of the possibilities to rewrite the Einstein equations as an
initial value problem for spatial hypersurfaces.  The dynamical fields
of the ADM formulation are a 3-metric $g_{ab}$ and its extrinsic
curvature $K_{ab}$ on a 3-manifold $\Sigma$, both depending on space
(points in $\Sigma$) and a time parameter, $t$. The foliation of the
4-dimensional spacetime into hypersurfaces $\Sigma$ is characterized
in the usual way by a lapse function $\alpha$ and a shift vector
$\beta^a$. The Einstein equations for vacuum become
\begin{eqnarray}
        (\partial_t - {\cal L}_\beta) \, g_{ab} &=& -2\alpha K_{ab} 
\label{dgdt}
\\
        (\partial_t - {\cal L}_\beta) \, K_{ab} &=& 
                              -D_aD_b\alpha + \alpha (R_{ab} 
                              - 2 K_{ac}{K^c}_b + K_{ab} K) 
\label{dKdt}
\\
        0 &=& D^b (K_{ab} - g_{ab} K) \equiv {\cal D}_a,
\label{diffeo}
\\
        0 &=& R - K_{ab} K^{ab} + K^2 \equiv {\cal H},
\label{hamil}
\end{eqnarray} 
where $R_{ab}$ is the 3-Ricci tensor, $R$ the Ricci scalar, $K$
the trace of the extrinsic curvature, ${\cal L}_\beta$ the Lie
derivative for $\beta$, and $D_a$ the covariant
derivative compatible with the 3-metric. One obtains evolution
equations for the metric variables, (\ref{dgdt}) and (\ref{dKdt}), and 
constraint equations that do not contain time derivatives of $g_{ab}$
or $K_{ab}$, the momentum constraint (\ref{diffeo}) and the
Hamiltonian constraint (\ref{hamil}). 

These equations are well known, but displaying them explicitly allows
me to make a number of basic observations. First of all, these are
comparatively simple equations. Even though writing out all the terms
in the index contractions, in the definition of $R_{ab}$ and the
covariant derivative leads to on the order of 1000 floating point
operations per point for a typical finite difference representation of
(\ref{dgdt}) and (\ref{dKdt}), this can be easily dealt with
computationally and is not one of the fundamental problems of black hole
evolutions. Still, a numerical implementation requires some thought
and hard work, see Sec.\ \ref{cactus}.

Notice that lapse and shift appear in the evolution equations
(as of course they have to) and have to be specified as part of
the evolution problem. Choosing lapse and shift fixes the coordinate
gauge for the evolutions, and is one of the key problems for numerical 
evolutions, see Sec.\ \ref{slicing}.

The constraint equations imply that specifying initial data $g_{ab}$
and $K_{ab}$ on a hypersurface $\Sigma$ involves in general solving the
constraints numerically. If the constraints are satisfied initially,
they will remain satisfied for a well-posed evolution system, but this is an
analytic statement that is only approximately true numerically.

Finally, the ADM equations do not define a hyperbolic evolution system
(e.g.\ \cite{Friedrich96}), and it is not clear to what extent the
original ADM equations can lead to a numerically stable evolution
system.  The issue of stability has to be addressed on two levels.
First, a well-posed evolution system is one for which existence,
uniqueness and stability of a solution for at least finite time
intervals can be shown, which is true for example for hyperbolic
systems. However, in general stability does not rule out exponentially
growing modes (this may be the solution one is looking for). Second,
the numerical implentation of an analytically stable system does not
trivially lead to stable numerical evolutions (e.g.\ the
finite-differenced equations may have exponentially growing solutions
which are not present analytically). Also note that important
stability issues arise at the boundaries of the computational domain.

Finding stable evolution systems is perhaps the other key issue in
numerical relativity besides the choice of coordinate problem.  For an
excellent review of first order hyperbolic systems for relativity see
\cite{Reula98a}, but other systems are of interest, too. One of the
important developments of the last year was the demonstration by
Baumgarte and Shapiro \cite{Baumgarte99}, that a conformal,
trace-split version of the ADM system very much like the system used
by Shibata and Nakumara in \cite{Shibata95}, is significantly more
stable (numerically) than the ADM equations for weak fields and some
algebraic slicings.  This BSSN system can also be understood as a
second order, conformal version of the Bona-Mass\'o system
\cite{Arbona99}. First order hyperbolic versions were given in
\cite{Alcubierre99c,Frittelli99}, although the BSSN system as it
stands is not hyperbolic. The BSSN variables are
\begin{eqnarray}
  \phi &=& \ln(\mbox{det} g)/12, \\
  K &=& g^{ab} K_{ab}, \\
  \tilde g_{ab} &=& e^{-4\phi} g_{ab}, \\
  \tilde A_{ab} &=& e^{-4\phi} (K_{ab} - g_{ab} K/3), \\
  \tilde \Gamma^c &=& \tilde \Gamma^c_{ab} \tilde g^{ab}, 
\end{eqnarray}
so that $\mbox{det} \tilde g =1$ and $\mbox{tr}\tilde A_{ab} = 0$.
Furthermore, introducing $\tilde \Gamma^c$ leads on the right-hand-side
of equation (\ref{dKdt}) to an elliptic expression in derivatives of
$\tilde g_{ab}$, i.e. the corresponding BSSN equation has the
character of a wave equation. However, including the evolution
equation for $\tilde \Gamma^c$ appears to spoil hyperbolicity.
Nevertheless, the BSSN system has very nice stability properties, and
some suggestions about why this may be the case are made in
\cite{Alcubierre99e}. Several BSSN evolutions have now been reported,
for strong waves and maximal slicing \cite{Alcubierre99b} (Sec.\
\ref{waves}) and also for matter evolutions
\cite{Baumgarte99b,Baumgarte99c,Alcubierre99d}.

\subsection{Schwarzschild as an example for a typical black hole 
evolution problem}
\label{slicing}

\begin{figure}
\centerline{\resizebox{6cm}{!}{\includegraphics{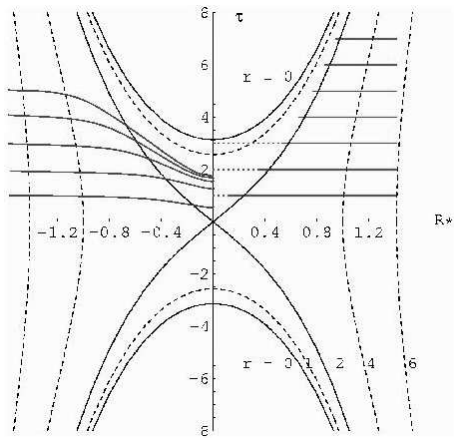}}}
 \caption{Schwarzschild black hole in Novikov coordinates, $M=1$.}
 \label{ssnovikov}
\end{figure}

Moving on to the prototypical example for a black hole evolution,
consider Fig.\  \ref{ssnovikov}, which shows the Schwarzschild
spacetime for a static, spherically symmetric black hole in Novikov
coordinates \cite{Novikov62,Misner73}. The coordinates are chosen such
that freely falling observers that start at rest at time $\tau = 0$
follow constant $R^*$ lines. The Schwarzschild radius $r$ is related
to $R^*$ at $\tau=0$ through $R^* = (r/(2M)-1)^{1/2}$. Several
constant $r$ lines are shown: the physical singularity at $r=0$, the
event horizon at $r=2M$, and note how the lines $r=4M$ and $r=6M$
curve outwards which corresponds to the radial infall of the observers
with constant $R^*$.

To set up an evolution problem we can choose the slice $\tau=0$ as
initial hypersurface with $g_{ab}$ and $K_{ab}$ derived from the
Schwarzschild four-metric ($g_{ab}$ and $K_{ab}$ therefore solve the
constraints). Note that the physical singularity is to the future of
this slice and does not show in $g_{ab}$ and $K_{ab}$. To perform an
evolution, we have to specify lapse and shift. The Novikov coordinates
correspond to geodesic slicing, $\alpha=1$, and vanishing shift,
$\beta^a=0$. Concretely, consider a numerical grid at $\tau=0$
extending from $R^*=0$ to $R^*=2^{1/2}$. In the first quadrant of the
figure we have shown how this initial slice moves through the
spacetime for geodesic slicing with vanishing shift. Without
precaution a numerical code will crash at $\tau = \pi M$ when the
point at $R^*=0$ reaches the singularity (this ``crash test'' has in
fact been used as a first crude code test \cite{Anninos94c,Bruegmann96}).

As shown in the figure, one can imagine evolving beyond $\tau = \pi M$
by cutting out from the slice what is inside the event horizon of the
black hole, which will not affect the outside of the black hole
anyway. ``Black hole excision'' techniques
\cite{Thornburg87,Seidel92a} are a very promising approach to black
hole evolutions, although in 3+1 dimensions there remain certain
stability problems to be resolved for binary black holes. Black hole
excision usually involves a non-trivial choice of lapse and shift.

With black hole excision not quite ready yet, it is so-called
singularity avoiding slicings that have been most widely used. Assume
$\beta^a=0$.  Primary examples are maximal slicing ($K=0$ initially
and $\Delta\alpha=\alpha K_{ab} K^{ab}$ so that $\partial_tK=0$), and
so-called ``1+log'' slicings (e.g.\ $\partial_t\alpha = - \alpha K/2$).
In the second quadrant of Fig.\ \ref{ssnovikov}, we show a typical
example (hand-drawn, while the rest of the figure was computed). At
the center, evolution slows down, while for large radii the evolution
marches on with $\alpha = 1$ at infinity. Obviously, a numerical
problem will occur in between, which is referred to as
grid-stretching, and which is reflected in growing sharp peaks in the
radial-radial component of the metric. Singularity avoiding slicings
have this fatal problem built in, but they allow us to compute
evolutions up to $30M-100M$, which is barely sufficient for certain
black hole collision and ring-down wave forms.

For completeness let me also mention the possibility of using
hyperboloidal slices, or null-slices, or null-slices matched to the
spatial slices, which when applicable cover more of the interesting
space time in the wave region, e.g.\ 
\cite{Friedrich81a,Friedrich81b,Huebner98,Frauendiener98c,Isaacson83,Bishop90,Winicour98,Winicour99}.
Characteristic matching is also useful near an excision boundary.

The key point to note is that the choice of coordinates in relativity
is a more fundamental problem than, say, the choice of spherical over
Cartesian coordinates for computational convenience. There is no
simple canonical choice like $\alpha=1$ and $\beta^a=0$ that works in
reasonably general situations. Even if there are no black holes,
geodesic slicing fails due to geodesic focusing. It appears to be the
case that one has to determine lapse and shift dynamically during the
evolution by some geometric principle as functions of the metric and
extrinsic curvature.

\subsection{Anatomy of a black hole simulation}
\label{skeleton}

After discussing 3+1 formulations in general and a specific black
hole example, let us list the components of a numerical relativity
evolution, with the binary black hole problem as example.

\subsubsection{Initial Data}
\begin{itemize}
\item Choice of hypersurface \\
  Simplest choice is $R^3$ for non-black hole data. Black hole data
  can be, for example, of Misner type with an isometry boundary
  condition at spheres representing the throats of the holes,
  $R^3-spheres$ \cite{Cook93}, or Brill-Lindquist type data based on a
  punctured $R^3$, $R^3-points$ \cite{Brandt97b}.
\item Solution to constraints \\
  There are four constraint equations that restrict the choice of 12
  components in $g_{ab}$ and $K_{ab}$. The most common approach is the
  conformal method \cite{Lichnerowicz44,Choquet62,York79}.
\end{itemize}

\subsubsection{Evolution}
\begin{itemize}
\item Variables and evolution system \\
  There are many different choices that can be roughly divided into
  ADM like systems that are of second order
  \cite{Shibata95,Baumgarte99}, and first order
  systems that are constructed to obtain hyperbolicity, e.g.\ 
  \cite{Reula98a}.  
\item Choice of coordinates (gauge choice) \\
  Typically a vanishing shift is used, but see Sec.\ \ref{shift}.
  For the lapse, as explained above, algebraic and elliptic conditions
  are in use.
\item Physical singularities \\
  Smooth, regular initial data may develop physical singularities
  which are features of black hole spacetimes. Physical singularities can
  be avoided by choice of slicing, or removed from the grid through
  black hole excision.
\item Coordinate singularities \\
  Dynamical determination of lapse and shift may lead to coordinate
  patho\-logies, in particular for algebraic slicings (e.g.\ 
  \cite{Alcubierre97a,Alcubierre97b}.  Elliptic conditions are
  sometimes preferable, although they are computationally much more
  expensive.
\item Outer boundary condition \\
  Asymptotic flatness can be assumed, which implies fall-off
  conditions for the fields. For run times on the order of $100M$ for
  a typical singularity avoiding Cauchy evolution, a radiative
  boundary condition is sufficiently accurate and stable, e.g.\ 
  \cite{Baumgarte99}. For a more sophisticated scheme see
  \cite{Abrahams97a}. Also there are two well developed approaches in
  which the numerical grid does not end at a finite radius but extends
  to future null infinity, either by matching to a characteristic code
  at finite radius \cite{Bishop90,Winicour98}, or smoothly without
  matching via a conformal transformation
  \cite{Friedrich81a,Friedrich81b,Huebner98,Frauendiener98a}. (Note
  that in 3+1 dimensions it is no longer straightforward to use a
  logarithmic radial coordinate as is conventionally done in
  axisymmetry.)
\item Inner boundary conditions \\
  As discussed above, black hole excision leads to a particular inner
  boundary. As for initial data construction, the inner boundary for
  slices in a black hole spacetime may be spherical or point-like.  For
  short term evolutions, the numerical slice can cover the inner
  asymptotically flat regions of the black holes if the resulting
  coordinate singularities are treated with the puncture method for
  evolution \cite{Anninos94c,Bruegmann97}.
\end{itemize}

\subsubsection{Analysis}
\begin{itemize}
\item Tensor components \\
  The raw output of a computer code will be the components of its
  basic variables, e.g.\ $g_{ab}$, $K_{ab}$, $\alpha$, and $\beta^a$,
  all other information is computed from these. Interesting local
  quantities include Riemann curvature invariants $I$ and $J$ and the
  Newman-Penrose invariants $\psi_0$ through $\psi_4$.
\item Black hole horizons \\
  The event horizon of black holes is a spacetime concept and can be
  found approximately if a sufficiently large spacetime slab has been
  computed, e.g.\ \cite{Matzner95a}. The apparent horizon is a notion
  intrinsic to the hypersurface (and is therefore slicing dependent).
  It is defined as the union of outermost marginally trapped surfaces,
  i.e. surfaces for which the expansion of outgoing null rays
  vanishes. Trapped surfaces are linked to the existence of black
  holes through the singularity theorems. See e.g.\ 
  \cite{Alcubierre98b} and references therein for numerical issues.
\item Wave extraction \\
  Wave forms can be computed reliably at finite but large radius using
  the first order gauge invariant approach of \cite{Abrahams88b}, as
  recently demonstrated for 3+1 dimensions in \cite{Allen98a}.
  In approaches that make future null infinity part of the numerical
  grid, see above, wave extraction is much more direct.  
\end{itemize}

\section{Implementation of a numerical relativity simulation}
\label{cactus}

As should be evident from the previous section, numerical relativity
poses a complex scientific problem that translates into a challenging
software engineering problem. Here I want to discuss ``Cactus'', a
code that is developed and used at the Albert-Einstein-Institut (AEI,
the Max-Planck-Institut f\"ur Gravitationsphysik), and several other
institutions \cite{Cactusweb,Seidel98c,Allen99a}.

Referring to \cite{Cactusweb}, the cactus code is a freely available
modular portable and manageable environment for collaboratively
developing high-performance multidimensional numerical simulations.
Cactus provides a powerful application programming interface based on
user modules (thorns) that plug into a compact core (flesh). Cactus is
composed of modules that are independent of relativity, and of modules
designed for relativity. The Cactus Computational Tool Kit supports a
variety of supercomputing architectures and clusters, implements
MPI-based parallelism for finite difference grids, several
input/output layers, elliptic solvers, metacomputing, distributed
computing, and visualization tools. Fixed and adaptive mesh refinement
is under development. Cactus significantly enhances collaborative
development by providing code sharing via CVS and defining appropriate
interfaces for code combination. A large number of physics modules or
thorns are available for numerical relativity and astrophysical
applications, e.g.\ there are thorns for initial data, evolution
routines, and data analysis. The first version of Cactus was created
by J. Mass\'o and P. Walker, and has been available for testing since
April, 1997 \cite{Masso98b}. The Cactus Computational Tool Kit (Cactus
4.0) saw its first public release as a community code in July, 1999.
It is actively supported by a cactus maintenance team, and there is
good documentation.  Cactus is a ``third generation'' code, going back
to the ``G'' \cite{Anninos94c} and ``H'' \cite{Anninos94d} codes. The
key step taken forward is the massive investment in the collaborative
infrastructure, which is now beginning to pay off. For many more
details, see \cite{Cactusweb,Seidel98c}.

So what does all of this mean? Suppose you want to run a black hole
simulation. Cactus is not a high-level science tool where you get an
executable with graphical user interface to input, say, the black hole
masses and off it goes. Numerical relativity is still too experimental
for that. At its heart, Cactus is a large collection of source files
together with a sophisticated make system. The user decides what
sources to include, then compiles the code. Runs are controlled by a
text file containing parameters, e.g.\ for the grid size and the black
hole masses. The sophistication lies in the ease with which code can be
changed or added by single users without affecting functionality
provided by others. Suppose a users wants to add a routine that
computes the determinant of the metric. A new thorn is created with
the source code, e.g.\ 20 lines of C or Fortran, and with files that
inform Cactus about new parameters, new grid functions (say an array
of reals with the name ``detg''), and tell cactus when to call the new
routine (in this case whenever analysis is done). This is work to be
done by the thorn writer, but he or she gets the rest for free: set-up
of a numerical grid, storage for $g_{ab}$ and its determinant,
evolution of $g_{ab}$ according to, say, the ADM equations, parallel
execution, input, output, adaptive mesh refinement, etc. When
submitted to the Cactus code repository, any Cactus user can now make
use of ``detg''.

Taking the viewpoint of a physicist, the Cactus infrastructure takes
care of many computer tasks that often distract from science. To say
that a simulation was carried out with Cactus can refer to Cactus, the
Computational Tool Kit, in the same way that credit is given to MPI
for parallelism, or Mathematica or Maple for symbolic computation.
Cactus is successful if the science outweighs the infrastructure.  In
order that Cactus does not remain a faceless collection of source
code, I would like to give several science examples and also to
mention at least a few names in connection with science projects.
Work on hyperbolic methods in numerical relativity \cite{Masso98a} was
done by J. Mass\'o, P. Walker, and others. A project on black hole
excision techniques has been implemented as ``Agave'' by S. Brandt, M.
Huq, P. Laguna, and others at Penn State University, which uses Cactus
mainly for parallelism.  Furthermore, the NASA Neutron Star Grand
Challenge Project of W.-M. Suen, E. Seidel, and others, develops the
so-called GR3D code \cite{Nasa98}, which is a version of Cactus for
coupled spacetime and relativistic hydrodynamics evolution based on
Riemann solvers. M. Miller implemented the key science module (MAHC)
for GR3D, with further contributions from other members of the Grand
Challenge \cite{Font98b,Miller99a}, see Sec.\ \ref{shift} for an
application.  Finally, Cactus is of course our platform for the binary
black hole collisions reported on in Sec.\ \ref{current} and the
strong gravitational wave evolutions of Sec.\ \ref{waves}. In this
case, the code ``BAM'' originally developed in
\cite{Bruegmann96,Bruegmann97a,Bruegmann97} contributed the multigrid
elliptic solver for the initial data and for maximal slicing, and the
Mathematica scripts of BAM were used to generate C code for the BSSN
evolution. For analysis, an apparent horizon finder implemented by M.
Alcubierre was used (there is also one available by C. Gundlach,
\cite{Alcubierre98b}), and the wave extraction routines by G. Allen
\cite{Allen98a}.  A much larger number of individuals than is apparent
from the above citations has contributed to ``the'' Cactus code, see
\cite{Cactusweb}. At this moment the Cactus 3.2 CVS repository lists
88 thorns, which range from private and under development to stable
and public.

\section{Grazing collision of black holes}
\label{current}

The first crude but truly (3+1)-dimensional binary black hole
simulation can be summarized as follows \cite{Bruegmann97}. The
approach taken was to address each of the items listed in the skeleton
for evolution problems of black holes of Sec.\ \ref{skeleton} in the
simplest possible manner that still allowed us to combine all the
ingredients to a complete implementation. Initial data for two black
holes, each with linear momentum and spin, is constructed using the
puncture method \cite{Brandt97b} (see also \cite{Beig96}), in which the
internal asymptotically flat regions of the holes are compactified so
that the numerical domain becomes $R^3$.  By construction the initial
data is conformally flat.  The evolution is performed with the
original ADM equations and a leapfrog finite difference scheme.
Maximal slicing and vanishing shift is chosen, i.e.\ physical
singularities are avoided, while coordinate singularities typically do
not occur for this elliptic slicing condition. At the outer boundary,
the ADM variables are held constant, which works well for the
achievable run times because a fixed mesh refinement of nested boxes
(with finer resolution at the center) is used, and for large radii the
lapse can approximate the Schwarzschild lapse for which Schwarzschild
data would remain static. An important insight is that the puncture
method, which can be made rigorous for initial data, can be
numerically extended to the evolution equations so that no special
inner boundary is present \cite{Bruegmann97}. Analysis is restricted
to apparent horizon finding with a curvature flow method.  These
methods allow one to evolve for about $7M$, which is sufficient to
observe the merger of the apparent horizons, but too short for wave
extraction.

Currently, binary black hole mergers are simulated by our
AEI/\-NCSA/\-WashU/\-Palma collaboration, and in this section I
summarize some of the preliminary results. These simulations build on
\cite{Anninos94c,Anninos96c,Bruegmann97} and introduce various
improvements. On the technical side, for high-performance
collaborative computing the code is implemented with Cactus 3.2. An
improved apparent horizon finder is now available
\cite{Alcubierre98b}. The comparatively slow maximal slicing can in
many situations be replaced by ``1+log'' slicing (cmp.\ Sec.\ 
\ref{slicing}). No mesh refinement is used, but the outer boundary is
treated with a radiative (Sommerfeld) boundary condition.  The above
plus the BSSN evolution system as given in \cite{Baumgarte99} with a
3-step iterative Crank-Nicholson (ICN) scheme, allow run times of up
to $30M$ for grazing collisions, compared to $7M$ for previous runs,
and up to $50M$ for simpler data sets.

The important new result is that now for the first time the extraction
of wave forms becomes possible with the methods tested in
\cite{Allen98a}. Let us discuss a concrete example. For initial data
we choose the punctures of each hole on the $y$-axis at $\pm 1.5$,
masses $m_1 =1.5$ and $m_2=1$, linear momenta $P_{1,2} = (\pm 2,0,0)$,
and spins $S_1=(-1/2,1/2,0)$ and $S_2=(0,1,1)$ (all units normalized
by $m_2=1$). The numerical grid has $385^3$ points with grid spacing
$0.2$, which puts the outer boundary for a centered cube at a
coordinate value of about $38$. The initial ADM mass is $M=3.11$, so
the outer boundary is at about $12M$ (solving the constraints for the
``bare'' parameters increases the mass over the Brill-Lindquist
vanishing spins and momenta value of $m_1+m_2$).  The total angular
momentum is $J = 6.7$, which corresponds to an angular momentum
parameter of $a/M = J/M^2 = 0.70$.

\begin{figure}
\centerline{\resizebox{12cm}{!}{\includegraphics{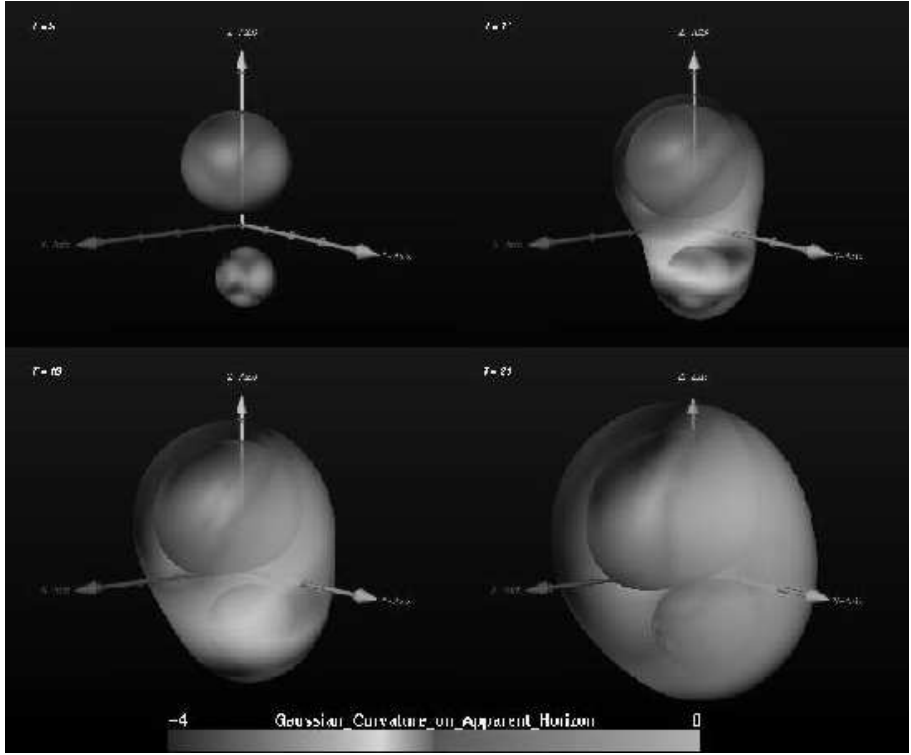}}}
 \caption{The evolution of the apparent horizon during a grazing black 
   hole collision ($t$ = 9, 11, 16, 21).}
 \label{blackholes}
\end{figure}

\begin{figure}
\centerline{
  \resizebox{6cm}{!}{\includegraphics{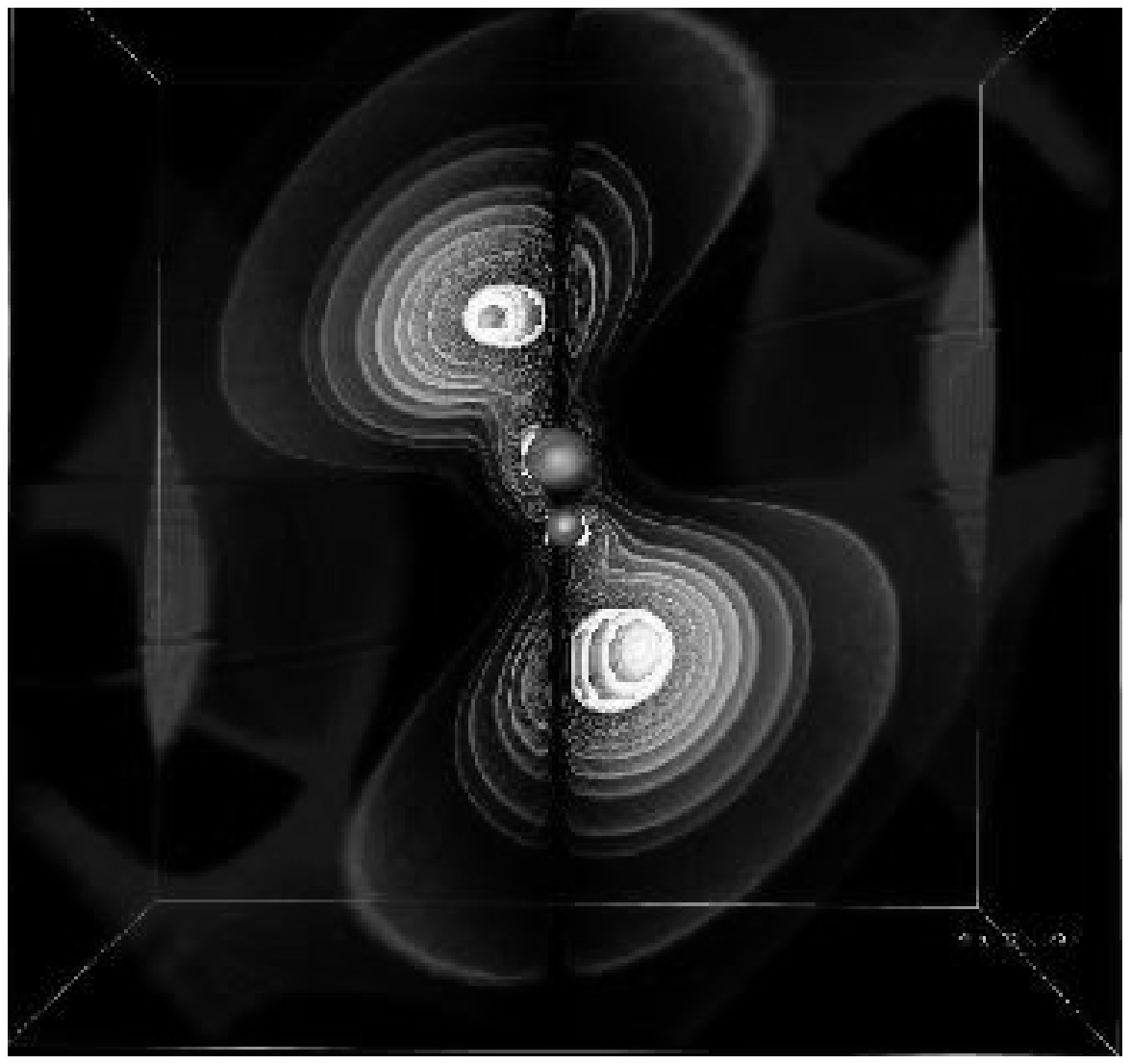}}
  \resizebox{6cm}{!}{\includegraphics{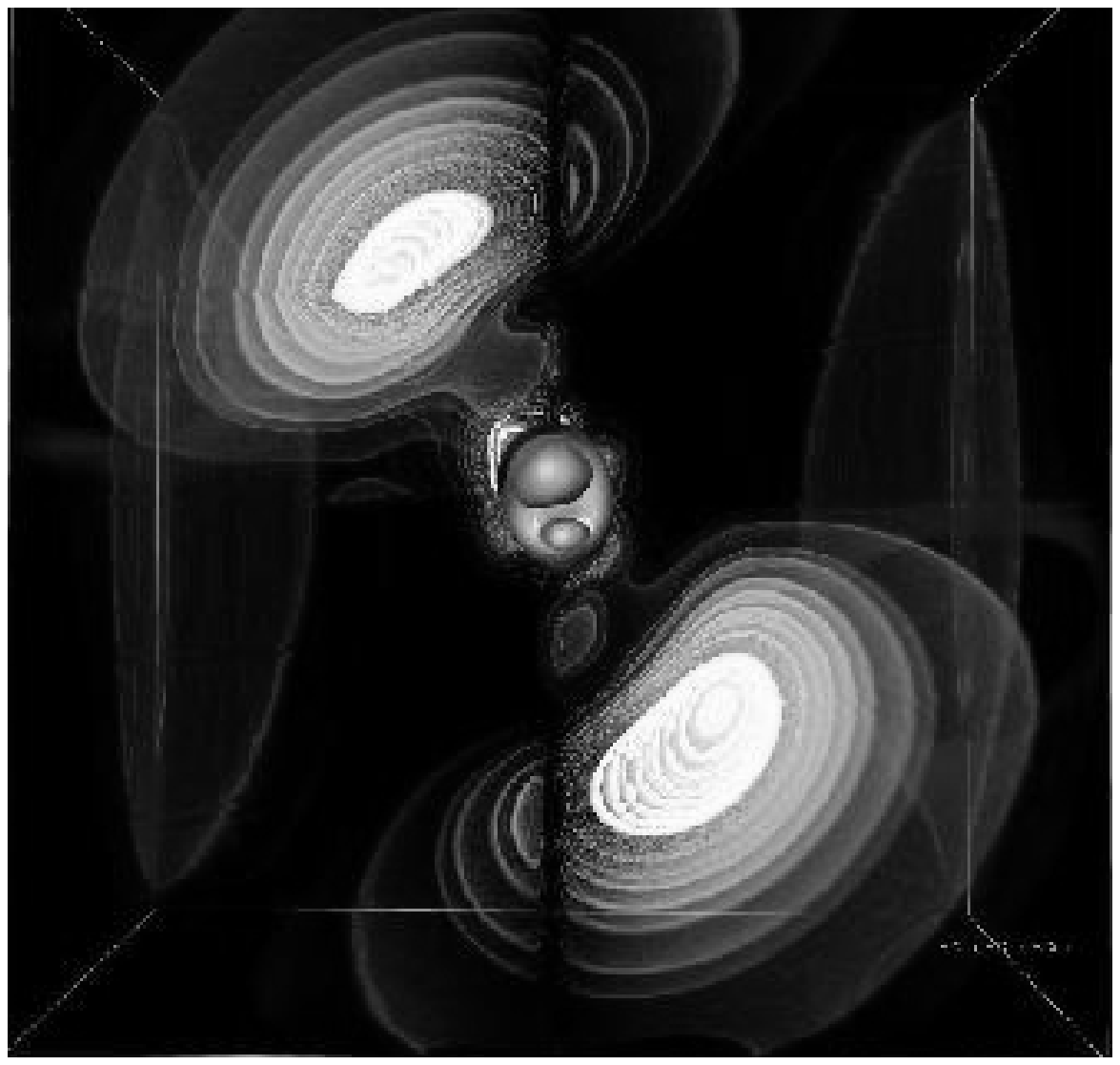}}}
 \caption{Apparent horizon during a grazing black hole collision with 
   Re$\psi_4$ as a wave indicator.}
 \label{blackholesandpsi4}
\end{figure}

The black holes start out with separate marginally trapped surfaces
forming the apparent horizon (although it may well be that they have a
common event horizon). Fig.\ \ref{blackholes} shows the formation of a
single marginally trapped surface surrounding the initial inner
marginally trapped surfaces. The apparent horizon is defined by a type
of minimal surface equation (e.g.\ \cite{Alcubierre98b}), and does not
evolve continuously, rather a new ``minimal'' surface appears in a new
location. The shading of the surfaces indicates the Gauss curvature on
the surfaces. The area of the apparent horizon increases in these
coordinates because grid points are falling into the black hole.
In Fig.\ \ref{blackholesandpsi4}, two frames near the
merger are shown together with isosurfaces of Re$\psi_4$ as a wave
indicator.

From the evolution, we can obtain an energy balance by comparing the
energy carried by the various modes of the gravitational waves
\cite{Allen98a} with the difference in mass between the initial slice
and the final black hole. For the latter, the apparent horizon mass is
$M_{AH} = [(M_{ir})^2 + J^2/(2M_{ir})^2]^{1/2}$ with $(M_{ir})^2 =
A_{AH}/(16\pi)$ and $A_{AH}$ the numerically determined area of the
apparent horizon of the final black hole. During the evolution,
$A_{AH}$ reaches a plateau, but then starts drifting upwards as the
grid stretching becomes more severe. With the plateau value for
$A_{AH}$ and $M_{ADM}\equiv M=3.11$ we obtain
\begin{equation}
   M_{ADM} - M_{AH} = 0.03 \approx 0.01 M_{ADM}.
\end{equation}
A rough estimate for the radiated energy in all modes until $t=30M$,
with the extraction radius rather close to the system at $8M$, is
\begin{equation}
   M_{RAD} = \mbox{0.007 -- 0.008} M_{ADM}.
\end{equation}
Even with all the current restrictions on accuracy coming from resolution,
grid size, boundary treatment, and grid stretching, this energy
balance can be considered to be a first physics result for such a
grazing collision. We learn that for this data set roughly 1\% of the
total energy is emitted in gravitational waves. Clearly, a thorough
parameter space study of such configurations is of interest. To
make contact with astrophysical situations, more realistic initial
data is probably needed (which ideally would be derived from the slow
inspiral of the two black holes). A detailed report is in preparation.

\section{Gravitational collapse of gravitational waves}
\label{waves}

One way to probe general relativity in the highly non-linear regime,
which should also share some of the strong wave features of the
grazing collision, is certainly through the gravitational collapse of
gravitational waves to a black hole. As briefly mentioned in Sec.\
\ref{2+1}, one scenario is that of critical collapse \cite{Choptuik93}.
One can construct a one-parameter family of initial data, and examine
the region near the ``critical'' value for that parameter at which a
black hole does or does not form. Not much is known in 3+1
dimensions \cite{Gundlach97d}, and the only study in axisymmetry is that
of Abrahams and Evans \cite{Abrahams92b,Abrahams93a} for gravitational waves.

In this section I want to briefly discuss first results for
non-axisymmetric collapse, cmp.\ \cite{Alcubierre99b}. We take as
initial data a pure Brill type gravitational wave \cite{Brill59},
later studied by Eppley~\cite{Eppley77,Eppley79} and
others~\cite{Holz93}.  The metric takes the form
\begin{equation}
ds^2 = \Psi^4 \left[ e^{2q} \left( d\rho^2 + dz^2 \right) 
+ \rho^2 d\phi^2 \right] =\Psi^4 \hat{ds}^{2},
\label{eqn:brillmetric}
\end{equation}
where $q$ is a free function subject to certain boundary conditions.
Following~\cite{Allen98a,Camarda97a,Brandt97c}, we choose $q$ of the
form
\begin{equation}
q = a \; \rho^2 \; e^{-r^2} \left[1 + c \; \frac{\rho^{2}}{(1+\rho^{2})} \;
\cos^{2} \left( n \phi \right) \right],
\end{equation}
where $a,c$ are constants ($a$ different from Sec.\ \ref{current}),
$r^2 = \rho^2 + z^2$ and $n$ is an integer.  For $c=0$, these data
sets reduce to the Holz~\cite{Holz93} axisymmetric form, recently
studied in three-dimensional Cartesian coordinates in preparation for
the present work~\cite{Alcubierre98b}.  Taking this form for $q$, we
impose the condition of time-symmetry, and solve the Hamiltonian
constraint numerically in Cartesian coordinates.  An initial data set
is thus characterized only by the parameters $(a,c,n)$.  For the case
$(a,0,0)$, we found in~\cite{Alcubierre98b} that no apparent horizon exists in
initial data for $a < 11.8$, and we also studied the appearance of an
apparent horizon for other values of $c$ and $n$.

For evolutions, we found that the BSSN system as given in
\cite{Baumgarte99} with maximal slicing, a 3-step ICN scheme, and a
radiative boundary condition is sufficiently reliable even for the
strong waves considered here. The key new extensions to previous BSSN
results are that the stability can be extended to (i) strong,
dynamical fields and (ii) maximal slicing, where the latter requires
some care.  Maximal slicing is defined by vanishing of the mean
extrinsic curvature, $K$=0, and the BSSN formulation allowed us to
cleanly implement this feature numerically, in contrast with the
standard ADM equations.

As discussed in \cite{Alcubierre99b}, axisymmetric data with $a=4$ is
subcritical, that is the imploding part of the wave disperses again,
leaving flat space in a non-trivially distorted coordinate system. An
amplitude of $a=6$ gives a supercritical evolution as indicated by the
formation of an apparent horizon. The ``cartoon'' method
\cite{Alcubierre99a} to perform axisymmetric calculations in Cactus
using three-dimensional Cartesian stencils on a two-dimensional slab
allowed us to close in on the critical region near $a=4.6$, but work
on detection of critical phenomena is still in progress.

\begin{figure} 
\vspace{-2mm}
\centerline{\resizebox{3.4in}{3.4in}{\includegraphics{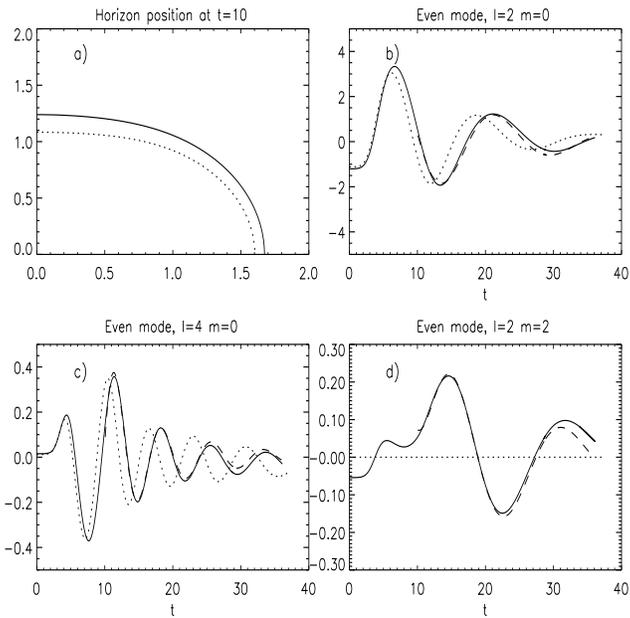}}}
\caption{
  a) The solid (dotted) line is the apparent horizon for the 3D data
  set $(6,0.2,1)$ ($(6,0,0)$) at $t$=10 on the $x$-$z$ plane.  b) The
  \{$l$=2,$m$=0\} wave form for the 3D $(6,0.2,1)$ case at $r=4$
  (solid line) is compared to axisymmetric $(6,0,0)$ case (dotted
  line). The dashed line shows the fit of the 3D case to the
  corresponding mode for a black hole of mass 1.0.  c) Same comparison for the
  \{$l$=4,$m$=0\} wave form. d) Same comparison for the
  non-axisymmetric \{$l$=2,$m$=2\} wave form.}
\label{fig:3D}
\end{figure}

Fig.~\ref{fig:3D} shows the development of the data set ($a$=6,
$c$=0.2, $n$=1), which has reflection symmetry across coordinate
planes. The initial ADM mass of this data set turns out to be
$M_{ADM}=1.12$.  Fig.~\ref{fig:3D}a shows a comparison of the apparent
horizons of this three-dimensional and the previous axisymmetric cases
at $t$=10 on the $x$-$z$ plane.  The mass of the three-dimensional
apparent horizon case is larger, weighing in at $M_{AH}$=0.99
(compared to $M_{AH}(2D)=0.87$).

In Fig.~\ref{fig:3D}b we show the \{$l$=2,$m$=0\} wave form of this
three-dimensional case, compared to the previous axisymmetric case.
The $c=0.2$ wave form has a longer wave length at late times,
consistent with the fact that a larger mass black hole is formed in
the three-dimensional case.  Figs.~\ref{fig:3D}c and~\ref{fig:3D}d
show the same comparison for the \{$l$=4,$m$=0\} and \{$l$=2,$m$=2\}
modes respectively.  Notice that while the first two modes are of
similar amplitude for both runs, the three-dimensional \{$l$=2,$m$=2\}
mode is completely different; as a non-axisymmetric contribution, it
is absent in the axisymmetric run (in fact, it does not quite vanish
due to numerical error, but it remains of order $10^{-6}$).  We also
show a fit to the corresponding quasi normal modes of a black hole of
mass 1.0.  The fit was performed in the time interval $(10,36)$, and
is noticeably worse if the fit is attempted to earlier times, showing
that the lowest quasi normal modes dominate at around $10$.  The early
parts of the wave forms $t<10$ reflect the details of the initial data
and BH formation process.  This is especially clear in the
\{$l$=2,$m$=2\} mode, which seems to provide the most information
about the initial data and the three-dimensional black hole formation
process.

\section{Minimal distortion shift}
\label{shift}

As a final example for recent advances in numerical relativity
simulations, let me mention shift conditions in (3+1)-dimensional
relativity. The first preliminary test of a dynamically computed
minimal distortion shift can be found in \cite{Daues96a} for a
Schwarzschild black hole on a 3+1 Cartesian grid, which is still the
only example with black hole excision. Computational domains with
holes pose a technical problem for the elliptic solver, which
certainly will be solved (see for example \cite{Diener99}) once
excision runs demand dynamic shifts.

A non-vanishing shift plays an important role in calculations that
involve orbiting black holes or neutron stars, e.g.\ in post-newtonian
calculations or Newtonian hydrodynamics for neutron stars. The freedom
in the shift vector can in principle be used to obtain corotating
coordinates or partially corotating coordinates (to counter frame
dragging). A variational principle to minimize coordinate shear leads
to the minimal distortion family of shift conditions, see
\cite{Smarr78b}. Introducing again a conformal factor such that the
conformal metric $\tilde g_{ab}$ has unit determinant, one can
minimize
\begin{equation}
S[\beta] = \int |{\partial_t} \tilde g|^2 dV = 
           \int \tilde g^{ac} \tilde g^{bd} 
                {\partial_t} \tilde g_{ab} {\partial_t} \tilde g_{cd} 
                \sqrt{\mbox{det} g} \, d^3x, 
\end{equation}
which gives a vector elliptic equation for $\beta^a$, 
\begin{eqnarray}
  (\Delta_l\beta)^a &=& 2 D_b(\alpha (K^{ab} - g^{ab} K/3)), \\
  ( \Delta_l\beta)^a & \equiv & D_bD^b\beta^a + D_bD^a\beta^b -
  \frac{2}{3} D^aD_b\beta^b.
\label{veclap}
\end{eqnarray}
Note that if there exists a rotational Killing vector, minimal
distortion can be trivially obtained \cite{Smarr78b}, hence such shift
conditions begin playing a non-trivial role only when one moves beyond
axisymmetric simulations (see also \cite{Beig97,Garfinkle99}, and in
particular \cite{Brady98a} for spacetimes with approximate Killing vector
fields).  

\begin{figure}
\centerline{\resizebox{10cm}{!}{\includegraphics{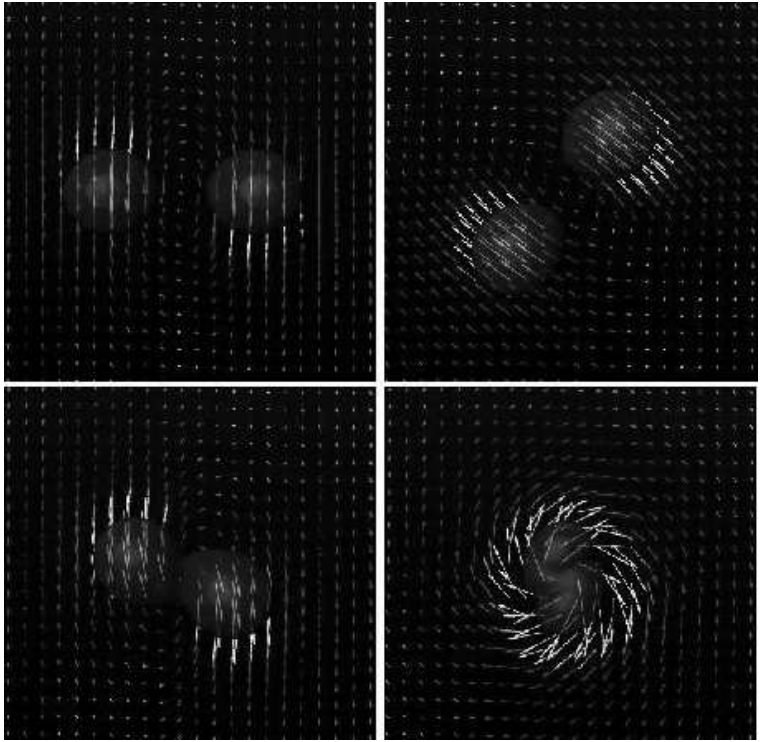}}}
 \caption{Binary neutron star orbit with minimal distortion shift. The 
   neutron stars are represented by an isosurface at 1/10 of the
   central density, the shift vector by the arrows.}
 \label{neutronstars}
\end{figure}

There are now three examples for the application of dynamical shift
conditions to binary neutron star simulations, which share the feature
that with vanishing shift the code fails after far less than an orbit,
while with minimal distortion shift for the first time fully
relativistic simulations of one or more orbits become possible. In
\cite{Nakamura99a,Shibata99a}, minimal distortion is approximated in a
way that decouples the three equations but maintains key features.
Preliminary experiments have also been performed for the NASA
Neutron Star Grand Challenge using Cactus, the hydrodynamics module or
``thorn'' MAHC \cite{Font98b,Miller99a}, and the author's
implementation of the vector Laplace operator (\ref{veclap}) in BAM.
The full minimal distortion equations are solved.  One choice of
initial data is that of irrotational neutron star binaries provided by
the Meudon group (\cite{Bonazzola98b,Bonazzola99a}, polytropic
equation of state with $\gamma=2$, $\kappa = 0.03 c^2 / \rho_{nuc} $,
$M_1=M_2=1.6M_{sol}$, $M/R=0.14$, $d=41km$). Fig.\  \ref{neutronstars}
shows four frames of an evolution project that was implemented and
carried out this summer by M. Miller, N. Stergioulas, and M. Tobias.
Without shift, the simulation crashes before less than 1/10th of an
orbit is completed, with shift one observes about 3/4 of an orbit
before the code fails when the two neutron stars merge. These first
results can probably be improved significantly, but they already serve
as a proof that non-vanishing shift is beneficial.

\section{Conclusion}
\label{conclusion}

It is perhaps surprising how little has been achieved to date by
numerical simulations of the Einstein equations for the two body
problem. After all, the Einstein equations have been extensively
studied for more than 80 years, and nowadays modern computational
physics has successfully treated the partial differential equations of
a large number of evolution problems. Why is it not possible to
``simply solve'' the problem with standard numerical methods on a big
computer? To recall some of the issues raised above, (i) the Einstein
equations do not lead to a unique or preferred set of 3+1 evolution
equations, with an automatically stable numerical implementation, (ii)
choosing lapse and shift is intricately coupled to the evolution,
(iii) black holes pose a special challenge due to their singularities.

As a result, black hole simulations in numerical relativity still have
to be called rather limited. Either special simplifications are
introduced (axisymmetry, null coordinates adapted to single black
holes), or the achieved numerical runtime is a limiting factor
compared to the lowest quasi-normal ringing period of about $17M$.
(3+1)-dimensional black hole evolutions with singularity avoiding
slicing last to about $30M$ for simple data sets starting from time
symmetry (vanishing extrinsic curvature) \cite{Anninos94c,Anninos96c}.
The first evolution of truly three-dimensional binary black hole data
(two black holes with spin and linear momentum) was performed in 1997
\cite{Bruegmann97}, crashing at $7M$, which allowed tracking the
merging of apparent horizons but not wave extraction. Considering that
the first 3+1 simulations of Schwarzschild were reported in 1995, one
can certainly call the recent simulations of Sec.\ \ref{current} with
wave extraction and a run time of about $30M$ a significant step
forward.

Several methods are under intense investigation that should allow us
to evolve for hundreds of $M$ or even longer. Here we mentioned black
hole excision, improved evolution schemes, and shift conditions.
Especially excision is expected to be essential. For the purpose of
wave extraction, the schemes involving future null infinity are of
particular interest. Furthermore, astrophysically more realistic
initial data is needed as input for the above methods before we can
make contact with gravitational wave astronomy.

How close is numerical relativity to the accurate prediction of
gravitational wave forms for binary events \cite{Goethe}?  The
post-newtonian and the close-limit approximations are probably in good
shape, but full numerical relativity will require two or more years to
get ready. An introductory statement often heard during the last two
decades is that one essential task of numerical relativity is to
provide a catalog of wave forms which is essential for gravitational
wave detection. This has changed. Numerical relativity will be
essential in wave analysis, producing models for astrophysical
scenarios that relate the wave forms to configuration parameters. For
the detection as such, however, the task of producing a complete
catalog appears to be too hard, and in particular, not a very sensible
one.  Note that matched filtering gives roughly a factor 5 in signal
to noise for wave detection \cite{Flanagan97a,Flanagan97b}. Recently,
the advantage of the perfect catalog over the best ``blind'' numerical
methods has been reduced to a factor of 2, e.g.\ \cite{Anderson99b}.
This still corresponds to a factor of 100 in observable event rate,
but on the other hand optimal matched filtering is assumed in this
estimate.  The emphasis in numerical relativity should therefore be
shifted more towards producing reliable statements about global
features of mergers as opposed to detailed wave forms.  Predicting the
duration of mergers, total energy emission, frequency range and
frequency distribution of the signal will be more useful to methods as
described in \cite{Flanagan97a,Flanagan97b,Anderson99b} and also more
attainable in the near future. The black hole runs of Sec.\ 
\ref{current} are being performed with this goal in mind.

\vspace*{0.25cm} \baselineskip=10pt{\small \noindent It is a pleasure
  to thank E. Seidel and all the members of the numerical relativity
  group at the AEI, and W.-M. Suen, M. Miller, and M. Tobias at WashU,
  St.\ Louis. Many colleagues have contributed without whom the recent
  work reported here would not have been possible. In particular, I
  would like to thank the Cactus support and development team, G.
  Allen, T. Goodale, G. Lanfermann, J. Mass\'o, M. Miller, and P.
  Walker, and in addition M. Alcubierre, S. Brandt, L. Nerger, E.
  Seidel, and R. Takahashi, with whom I have collaborated on the black
  hole runs reported in Sec.\ \ref{current}. Figs.\ \ref{blackholes},
  \ref{blackholesandpsi4}, and \ref{neutronstars} were prepared by W.
  Benger with the Amira software of ZIB, see
  \cite{Potsdam-WashU-movies}.  This work has been supported by the
  AEI, NCSA, NSF PHY 9600507, NSF MCA93S025 and NASA NCCS5-153.
  Calculations were performed at AEI, NCSA, RZG in Garching, and ZIB
  in Berlin.  }


\end{document}